\documentstyle[12pt]{article}
\input epsf

\newcommand{\be}{\begin{equation}}

\newcommand{\ee}{\end{equation}}
\newcommand{\bea}{\begin{eqnarray}}
\newcommand{\eea}{\end{eqnarray}}
\newtheorem{algo}{Algorithm}
\begin{document}
\begin{center}
{\LARGE {\bf Generalized Hyper-Systolic Algorithm}}\\[3cm]
{\Large {\bf A. Galli}}\\[1cm]
{Max-Planck-Institut f\"ur Physik, F\"ohringer Ring 6, D-80805 M\"unchen}
\footnote{e-mail: galli@mppmu.mpg.de}\\[3cm]
{\bf Abstract}
\end{center}
We generalize the hyper-systolic algorithm proposed
in \cite{schilling} for abstract data structures on massive parallel
computers with $n_p$ processors.
For a problem of size $V$ the communication complexity of the
hyper-systolic algorithm is proportional to $\sqrt{n_p}V$, to be
compared with $n_pV$ for the systolic case.
The implementation technique is explained in detail and the example
of the parallel matrix-matrix multiplication is tested on the Cray-T3D.
\newpage
\section{Introduction}
In a near future Teraflops supercomputers will be
available from many vendors. A possible architecture to achieve such a
high performance is based on massive parallelism. Many  powerful
processors (PE) are linked together. Each of them can access his own local
memory which is usually relatively large.
The communications between the PE's is performed by a fast network.
Using this architecture the hope is to obtain a linear speedup
in the number of PE's. This is,
of course, not possible because the overhead of the communications increases
with the number of PE's.\\

The overhead due to the communications can  be encountered in many
algebraic problems. For example, let us assume we have two matrices
$A$ and $B$ of size $V$ which are
distributed among the PE's, then the simple operation $A\cdot B$ needs
inter processor communications. The volume of the data which have to be
communicated is then proportional to $n_p$ keeping the size of the matrices
fixed. In fact the volume of private data owned by one PE is proportional to
$V/n_p$. Using an usual systolic algorithm \cite{sys}
the matrix-matrix multiplication needs that each PE communicates his
local portions of the matrices
to each PE so that the volume of the communicated data will be
$n_p^2V/n_p$ which is proportional to $n_p$. \\

The granularity of a
problem is considered to be the average size of tasks to be performed
in parallel.
A problem of size $V$ implemented in parallel on $n_p$ processors has
a granularity $g$ proportional to $V/n_p$.
If the granularity of the problem is very large the
inter processor communication becomes irrelevant compared with the
time spent for the local computations. But if the granularity of the problem
is small the overhead due to the communication
is a serious obstacle to the wanted speedup.
In fact, using an usual systolic algorithm, the data to be communicated is
proportional to $Vn_p$ and because we can
assume that each PE can communicate data in parallel, the overhead time is
constant in the number of PE's $Vn_p/n_p=V$.
The computer time needed to
solve a problem of granularity $g$ is then
proportional to $\tau=a\cdot \frac{V}{n_p}+b\cdot V$ where $a$ and $b$ are
constants. Asymptotically in $n_p$ this time reach a plateau given by the
overhead time $b\cdot V$.\\

Of course, performance can be obtained by increasing the size of the problem
so that the granularity remains constant. This is nice, but usually it is
not what one needs in real applications. In fact, one may also
needs speedup by more or less maintaining constant the size of the problem.
It is then an important task to find new algorithms that need less inter
processors communications even at low granularity.
This is possible exploiting the fact that usually
at low granularity there is abundance of memory. \\

Recently, it was proposed an {\em hyper-systolic} \cite{schilling}
algorithm for organizing the data communications related to the problem
\be
y_i=\sum_{j\mbox{ over all PE}} f(x_i,x_j) \label{1}
\ee
where $x_i$ and $y_i$ are some data stored on the $i$ PE and $f$ is some
function. It was shown that the volume of the communicated data was
proportional to $\sqrt{n_p}$ for fixed size of the problem\footnote{In the
original work of \cite{schilling} it is showed that the communicated data is
proportional to $n_p^{3/2}$ if one increase the size of the problem linearly in
$n_p$.}.
This algorithm can be useful in various fields of applications, like $n-$body
dynamics, polymer chains, protein folding or signal processing.\\

Equation (\ref{1}) describes a small class of problems. Of course,
the hyper-systolic algorithm can be generalized for a more
large class of problems with global communication and with a more
complicate data structure. We consider the following generalized problem
$$
C_{i_{mype}}= \bigoplus_{i_p\mbox{ over all
PE}}f(A_{i_{mype}},B_{i_p})
$$
where now $A,B$ and $C$ are abstract data structures, $f$ is some
operations on $A$ and $B$ which has an output with the same data structure
type of $C$ and $\bigoplus$ is an associative and commutative
operation on the data structure of type
$C$. The portions of the data owned by the different PE's are denoted by
the subscript $i_p$. The index $i_{mype}$ characterizes the local PE.\\

For example, if one
considers the matrix-matrix product the same formalism can be applied.
In this case $A_{i_{mype}}$ and $C_{i_{mype}}$ are the portions of the
matrices $A$ and $C$ stored locally on the PE $i_{mype}$ and
$B_{i_p}$ is the portion of the matrix $B$ stored on the remote PE $i_p$.
The product $f(A_{i_{mype}},B_{ip})=A_{i_{mype}}*B_{ip}$
is a generalized matrix product which takes care
how the matrices are distributed among the PE's
(column-like, row-like or block-like)
and the operator $\bigoplus$ is a direct sum on the respective $n_p$
linear subspaces of $C_{i_{mype}}$.\\

The hyper-systolic algorithm replicas a volume of data proportional to
$\sqrt{n_p}V/n_p$ coming from the portions of $A$ and $B$ stored on
$O(\sqrt{n_p})$ different PE's, and uses
all possible combinations of them to built $O(\sqrt{n_p})$ different
portions of $C$ which are then sent back to the original PE's.
The communication directions are chosen so that all PE's receive all needed
data. In this case the amount of data to be communicated is only
proportional to $V\sqrt{n_p}$ for a fixed size $V$ of the problem. Then the
time needed to solve a problem of granularity $g$ is
$\tau=a\cdot\frac{V}{n_p}+b\cdot \frac{V}{\sqrt{n_p}}$ where $a$ and $b$ are
some constants. Now the overhead time $b\cdot \frac{V}{\sqrt{n_p}}$ vanishes
asymptotically in the number of PE's.\\

In this paper we discuss in details the hyper-systolic algorithm
for an abstract data structure and then we consider
the useful example of the matrix-matrix product which
is one of the most hard problem in linear algebra operations
on massive parallel
computers and also mimics all aspects of more general applications.
We also report
the gain in performance against the usual systolic algorithm
tested on the Cray-T3D.

\section{Generalized Algorithm}

We consider a massive parallel computer built up by a set of
$n_p$ PE's labeled by $S_{PE}=\{i_p\in\{0,...,n_p-1\}\}$ and a
network connecting the PE's
$N=\{b_{i,j}|i,j\in S_{PE} \}$.
The index $i_{mype}$ denotes the local PE.
We define a set of three data structures $\hat D=\{\hat A,\hat B,\hat C\}$.
Each PE owns a portion of $\hat D$. We denote this portion by $\hat D_{i_p}$.
For three
variables $A\in \hat A, B\in \hat B$ and $C\in \hat C$ we want
to compute
\be
C_{i_{mype}}= \bigoplus_{i_p\mbox{ over all PE}}f(A_{i_{mype}},B_{i_p})
\label{master}
\ee
where $f:\{\hat A_{i_{mype}},\hat B_{i_{mype}}\}\rightarrow \hat
C_{i_{mype}}$
is some
operations on the local $\hat A$ and $\hat B$ which has an output of type
$\hat C$ and
$\bigoplus:\{\hat C_{i_{mype}},\hat C_{i_{mype}}\}\rightarrow
\hat C_{i_{mype}}$
is an operation with associative and commutative
properties on the data structure of type
$\hat C$ with output of type $\hat C$. \\

The usual systolic algorithm works as follow
\begin{algo}{Systolic Algorithm}
{\rm
\bea
\mbox{do}&&\forall i_p\in S_{PE}\nonumber\\
&& \mbox{get $B_{i_p}$ from PE $i_p$}\nonumber\\
&& C_{i_{mype}}=C_{i_{mype}}\oplus f(A_{i_{mype}},B_{ip})\nonumber\\
\mbox{od}&&\nonumber
\eea}
{\rm where the communication routine $get$
represents a shift sequence optimized for the used architecture. We do not
go into the details of how this routine organizes the shifts because it is
standard.
}
\end{algo}

The idea of the hyper-systolic algorithm is to allocate on each PE a set of
replicas $\{D^k_{i_{mype}} \in \hat D_{i_{mype}}|k\in I\}$.
The set of integer $I=\{0,...,K\}$ contains the index of the replicas of
$D_{i_{mype}}$ that we want to memorize on each PE.
Each replica $A^k$ and $B^k$ has to be filled with a copy of some
remote portion of $A$ and $B$. Then many combinations of these portions can be
used to find many contributions to $f(A_{i_{mype}},B_{ip})$.
These have then to be sent back to the right PE's. To control all
the directions of communications one has to ensure that all needed
contributions occur. This can be done by
defining a mapping $\phi$ from $S_{PE}\times I$ to $S_{PE}$
by the following algorithm
\begin{algo}
{\rm
\bea
\mbox{do}&&\forall i_p\in S_{PE}\nonumber\\
&&i_{tmp}=i_p\nonumber\\
&&\mbox{do}\, \forall k\in I\nonumber\\
&&\,\,\,\,\,\,\phi(i_p,k)=i_{tmp}-\Gamma_k\nonumber\\
&&\,\,\,\,\,\,\mbox{while }\phi(i_p,k)< 0\mbox{ do}\nonumber\\
&&\,\,\,\,\,\,\,\,\,\,
\phi(i_p,k)=\phi(i_p,k)+n_p\nonumber\\
&&\,\,\,\,\,\,\mbox{od }\nonumber\\
&&\,\,\,\,\,i_{tmp}=\phi(i_p,k)\nonumber\\
&&\mbox{od}\nonumber\\
\mbox{od}\nonumber
\eea}
\end{algo}
where the $\Gamma_k=\{0,\gamma_1,...,\gamma_K\}$ with $\gamma_i$
positive and integer. The dimension of the vector $\Gamma$
has to be chosen with $K$ minimal so that
\be
\forall m\in S_{PE}: m=\gamma_i+\gamma_{i+1}+...+\gamma_{i+j}\,\mbox{ or }
m=n_p-(\gamma_i+\gamma_{i+1}+...+\gamma_{i+j})\label{ak}
\ee
with $0\leq i+j\leq K$. How to determine the appropriate vector
$\Gamma$ is discussed in \cite{schilling}. The
best choice of the vector $\Gamma$ is called the best basis. The best
basis has dimension $K\simeq \sqrt{n_p}$. There is no analytical
recipes for finding the best basis for a given number of PE's. One has
to try all combinations of numbers and select the ones which satisfy
(\ref{ak}).
But this can be a very hard computational problem for a
large number of PE's.
For the allowed configurations of the Cray-T3D we have enumerate the
best basis in Table 1.
Fortunately, there exists a regular basis which is
very easy to be found and has dimension $K\simeq \sqrt{2\cdot n_p}$.
This basis  is given by
\be
\Gamma=(\underbrace{1,1,...,1}_{K},\underbrace{K,K,...,K}_{K-1})
\ee
where $K$ is the integer part of $\sqrt{2n_p}$.
In any case the choices of  $K$ are of order $O(\sqrt{n_p})$.\\

The mapping $\phi$ is local to each PE and it is equal for each PE. It
contains all informations about the location of the data $D$ which has to be
communicated to fill the $K$ replicas. The definition (\ref{ak}) ensures that
all combinations of PE's occurs at least once.
To ensure that only one PE will compute the combination $f(A^{k_1}_{i_{mype}}
,B^{k_2}_{i_{mype}})$ and send it back to the right PE one has to check the
multiplicity of the occurrence of this combination among the PE's
defining a multiplicity table $M$ going from
$I^2$ to $\{0,1\}$.
All informations
about the multiplicity of the operations can be read from the mapping $\phi$.
$M$ takes the value $0$ when a duplicate of the combination
$f(A^{k_1}_{i_{mype}}
,B^{k_2}_{i_{mype}})$ is encountered.
This table controls
that in the operation $\oplus$ the
elements contributes only once by switching off the PE's
which duplicates this operation when
$M(k_1,k_2)=0$.
\begin{algo}
{\rm
\bea
&&M(\phi(i_{mype},k_1),\phi(i_{mype},k_2))=1\,\,\forall k_1,k_2\in I\nonumber\\
\mbox{do}&& i_p=0,i_{mype}-1\nonumber\\
&&\mbox{do }\forall k_1,k_2,k_3,k_4\in I\nonumber\\
&&\,\,\,\,\,\mbox{if } (\phi(i_{mype},k_1)=\phi(i_p,k_3)\mbox{ and }
\phi(i_{mype},k_2)=\phi(i_p,k_3))\mbox{ then }
\nonumber\\
&&\,\,\,\,M(\phi(i_{mype},k_1),\phi(i_{mype},k_2))=0
\nonumber\\
&&\,\,\,\,\mbox{fi}\nonumber\\
&&\mbox{od}\nonumber\\
\mbox{od}&&\nonumber
\eea}
\end{algo}
The hyper-systolic algorithm can then be organized as following: one gets
from the remote PE's
selected by the $\phi$ mapping
the $K$ portions of data $\{A,B\}$
needed to built all $K^2$ combinations
$f(A^{k_1}_{i_{mype}},B^{k_2}_{i_{mype}})$ among the replicas.
Then one sums up these
combinations in the respective replica $C^{k_1}_{i_{mype}}$ according to the
operation $\oplus$ and controlling the multiplicity.
Now one has to synchronize\footnote{Of course, the synchronization is needed
only on MIMD architectures.} all PE's to be sure that all
have done their job. Then one has to send back according to the exact reverse
$\phi(i_{mype},k)$ sequence all $C^{k_1}_{i_{mype}}$ and sum up them to the
result local $C$ according with $\oplus$.
\begin{algo}{Hyper-Systolic Algorithm}
{\rm
\bea
\mbox{do}&&\forall k\in I\nonumber\\
&&\mbox{get $A_{i_p}$ from PE $i_p=\phi(i_{mype},k)$ and store in
$A^k_{i_{mype}}$}
\nonumber\\
&&\mbox{get $B_{i_p}$ from PE $i_p=\phi(i_{mype},k)$ and store in
$B^k_{i_{mype}}$}
\nonumber\\
\mbox{od}&&\nonumber\\
\mbox{do}&&\forall k_1,k_2\in I\nonumber\\
&&\mbox{if } M(k_1,k_2)=1 \mbox{then }
C^{k_1}_{i_{mype}}=C^{k_1}_
{i_{mype}}\oplus f(A^{k_1}_{i_{mype}},B^{k_2}_{i_{mype}})\nonumber\\
\mbox{od}&&\nonumber\\
&&\mbox{synchronize}\nonumber\\
\mbox{do}&&\forall i_p\in S_{PE},k\in I\nonumber\\
&&\mbox{if }\phi(i_p,k)=i_{mype} \mbox{then}\nonumber\\
&&\,\,\,\mbox{get $C^k_{i_p}$ from PE $i_p$ and store in $T\in\hat
C_{i_{mype}}$}
\nonumber\\
&&\,\,\,C_{i_{mype}}=C_{i_{mype}}\oplus T\nonumber\\
&&\mbox{fi}\nonumber\\
\mbox{od}&&\nonumber
\eea}
\end{algo}
This pseudo-code can be used to implement any parallel problem satisfying eq.
(\ref{master}).
As a test example we have implemented the product of
two matrices on the Cray-T3D.
We have distributed the matrices so that each PE owns
some fixed portion of rows or columns of the matrices.
We have tested the
algorithm for matrices with dimensions going from $16\times 16$ to
$2048\times 2048$ real entries on $2$ to $256$ PE's.
We have measured the ratio between the
time spent to communicate data for the systolic against the hyper-systolic
algorithm. This ratio expresses the gain in computer time for the
communications.
The hyper-systolic algorithm communicates three
times a volume of data equal to $K\cdot V$ (where $K$ is the integer
part of
$\sqrt{n_p}$ for the best basis or the integer part of $\sqrt{2n_p}$
for the regular basis)
and the systolic algorithm communicates a volume of
data equal to $n_p\cdot V$. In the ideal case we can assume that
the networks is able to transfer data in parallel with the same speed
for both algorithms then the ideal ratio is
\bea
&&R_{\mbox{ideal}}\simeq\frac{n_p}{3\cdot\sqrt{n_p}}=\frac{\sqrt{n_p}}{3}
\,\,\,\mbox{best basis}\nonumber\\
&&R_{\mbox{ideal}}\simeq\frac{n_p}{3\cdot\sqrt{2\cdot n_p}}=
\frac{\sqrt{n_p}}{3\sqrt{2}}\,\,\,\mbox{regular basis}
\eea
Of course, the hyper-systolic algorithm can not have an ideal ratio against the
systolic algorithm because it involves communications to more remote PE's
than in the systolic case. This eventually can overload the
network and slow down the communications. However,
if the topology of the network is a $d-$torus
with $d\geq 2$ this effect should be negligible because there
exist many paths which connect each pair of PE's. Only if
$d=1$ then a $1-$torus is a circle
some problem can arise because for connecting a pair of PE's there exist
only two paths.
In fig. 1 we plot the measured ratio averaged on the different
dimensions of the matrices.
We see that the measured values lies near the ideal case as expected
from the topology of the Cray-T3D.
\section{Conclusion}
We have generalized the hyper-systolic algorithm proposed by
\cite{schilling}. This algorithm is able to handle a vast class of
massive parallel problems which need global communications.
It reduces significantly the overhead due to the
communications when compared with the usual systolic algorithm. As an
example we have implemented the algebraic problem of multiplying two
large matrices in parallel. In our example we have observed
a speedup of the communications of about $\frac{\sqrt{n_p}}{3}$ using the best
basis and $\frac{\sqrt{n_p}}{3\sqrt{2}}$ using the regular basis.\\[1cm]

{\Large {\bf Acknowledgments}}\\

I would like to thank T. Lippert for discussions and help. I also
would like to thank N. Galli for helpful comments and discussions.

\begin{table}
\begin{center}
\begin{tabular}{|c|c|l|}\hline
$n_p$ & $K$ & $\Gamma$\\\hline\hline
2 & 1 & 1\\
4 & 2 & 1 1\\
8 & 3 & 1 1 2\\
16& 4 & 1 2 2 4\\
32& 6 & 1 1 1 4 4 8\\
64& 8 & 1 1 12 3 10 8 20 4\\
128& ?& ?\\
256& ?& ?\\\hline
\end{tabular}
\end{center}
\caption{The best basis for the configurations of the Cray-T3D at the EPFL,
Switzerland. The
basis for $128$ and $256$ PE's could not be found. }
\end{table}

\begin{figure}
\setlength{\unitlength}{0.240900pt}
\ifx\plotpoint\undefined\newsavebox{\plotpoint}\fi
\sbox{\plotpoint}{\rule[-0.200pt]{0.400pt}{0.400pt}}%
\begin{picture}(1500,900)(0,0)
\font\gnuplot=cmr10 at 10pt
\gnuplot
\sbox{\plotpoint}{\rule[-0.200pt]{0.400pt}{0.400pt}}%
\put(220.0,113.0){\rule[-0.200pt]{292.934pt}{0.400pt}}
\put(220.0,113.0){\rule[-0.200pt]{0.400pt}{184.048pt}}
\put(220.0,113.0){\rule[-0.200pt]{4.818pt}{0.400pt}}
\put(198,113){\makebox(0,0)[r]{0}}
\put(1416.0,113.0){\rule[-0.200pt]{4.818pt}{0.400pt}}
\put(220.0,240.0){\rule[-0.200pt]{4.818pt}{0.400pt}}
\put(198,240){\makebox(0,0)[r]{1}}
\put(1416.0,240.0){\rule[-0.200pt]{4.818pt}{0.400pt}}
\put(220.0,368.0){\rule[-0.200pt]{4.818pt}{0.400pt}}
\put(198,368){\makebox(0,0)[r]{2}}
\put(1416.0,368.0){\rule[-0.200pt]{4.818pt}{0.400pt}}
\put(220.0,495.0){\rule[-0.200pt]{4.818pt}{0.400pt}}
\put(198,495){\makebox(0,0)[r]{3}}
\put(1416.0,495.0){\rule[-0.200pt]{4.818pt}{0.400pt}}
\put(220.0,622.0){\rule[-0.200pt]{4.818pt}{0.400pt}}
\put(198,622){\makebox(0,0)[r]{4}}
\put(1416.0,622.0){\rule[-0.200pt]{4.818pt}{0.400pt}}
\put(220.0,750.0){\rule[-0.200pt]{4.818pt}{0.400pt}}
\put(198,750){\makebox(0,0)[r]{5}}
\put(1416.0,750.0){\rule[-0.200pt]{4.818pt}{0.400pt}}
\put(220.0,877.0){\rule[-0.200pt]{4.818pt}{0.400pt}}
\put(198,877){\makebox(0,0)[r]{6}}
\put(1416.0,877.0){\rule[-0.200pt]{4.818pt}{0.400pt}}
\put(220.0,113.0){\rule[-0.200pt]{0.400pt}{4.818pt}}
\put(220,68){\makebox(0,0){0}}
\put(220.0,857.0){\rule[-0.200pt]{0.400pt}{4.818pt}}
\put(423.0,113.0){\rule[-0.200pt]{0.400pt}{4.818pt}}
\put(423,68){\makebox(0,0){50}}
\put(423.0,857.0){\rule[-0.200pt]{0.400pt}{4.818pt}}
\put(625.0,113.0){\rule[-0.200pt]{0.400pt}{4.818pt}}
\put(625,68){\makebox(0,0){100}}
\put(625.0,857.0){\rule[-0.200pt]{0.400pt}{4.818pt}}
\put(828.0,113.0){\rule[-0.200pt]{0.400pt}{4.818pt}}
\put(828,68){\makebox(0,0){150}}
\put(828.0,857.0){\rule[-0.200pt]{0.400pt}{4.818pt}}
\put(1031.0,113.0){\rule[-0.200pt]{0.400pt}{4.818pt}}
\put(1031,68){\makebox(0,0){200}}
\put(1031.0,857.0){\rule[-0.200pt]{0.400pt}{4.818pt}}
\put(1233.0,113.0){\rule[-0.200pt]{0.400pt}{4.818pt}}
\put(1233,68){\makebox(0,0){250}}
\put(1233.0,857.0){\rule[-0.200pt]{0.400pt}{4.818pt}}
\put(1436.0,113.0){\rule[-0.200pt]{0.400pt}{4.818pt}}
\put(1436,68){\makebox(0,0){300}}
\put(1436.0,857.0){\rule[-0.200pt]{0.400pt}{4.818pt}}
\put(220.0,113.0){\rule[-0.200pt]{292.934pt}{0.400pt}}
\put(1436.0,113.0){\rule[-0.200pt]{0.400pt}{184.048pt}}
\put(220.0,877.0){\rule[-0.200pt]{292.934pt}{0.400pt}}
\put(45,495){\makebox(0,0){Ratio }}
\put(828,23){\makebox(0,0){$n_p$}}
\put(220.0,113.0){\rule[-0.200pt]{0.400pt}{184.048pt}}
\put(228,152){\raisebox{-.8pt}{\makebox(0,0){$\Diamond$}}}
\put(236,164){\raisebox{-.8pt}{\makebox(0,0){$\Diamond$}}}
\put(252,189){\raisebox{-.8pt}{\makebox(0,0){$\Diamond$}}}
\put(285,253){\raisebox{-.8pt}{\makebox(0,0){$\Diamond$}}}
\put(350,317){\raisebox{-.8pt}{\makebox(0,0){$\Diamond$}}}
\put(479,419){\raisebox{-.8pt}{\makebox(0,0){$\Diamond$}}}
\put(228,136){\makebox(0,0){$+$}}
\put(236,149){\makebox(0,0){$+$}}
\put(252,170){\makebox(0,0){$+$}}
\put(285,209){\makebox(0,0){$+$}}
\put(350,259){\makebox(0,0){$+$}}
\put(479,323){\makebox(0,0){$+$}}
\put(739,425){\makebox(0,0){$+$}}
\put(1258,571){\makebox(0,0){$+$}}
\sbox{\plotpoint}{\rule[-0.400pt]{0.800pt}{0.800pt}}%
\put(228,173){\usebox{\plotpoint}}
\multiput(229.40,173.00)(0.512,1.486){15}{\rule{0.123pt}{2.455pt}}
\multiput(226.34,173.00)(11.000,25.905){2}{\rule{0.800pt}{1.227pt}}
\multiput(240.40,204.00)(0.514,1.155){13}{\rule{0.124pt}{1.960pt}}
\multiput(237.34,204.00)(10.000,17.932){2}{\rule{0.800pt}{0.980pt}}
\multiput(250.40,226.00)(0.514,0.988){13}{\rule{0.124pt}{1.720pt}}
\multiput(247.34,226.00)(10.000,15.430){2}{\rule{0.800pt}{0.860pt}}
\multiput(260.40,245.00)(0.512,0.788){15}{\rule{0.123pt}{1.436pt}}
\multiput(257.34,245.00)(11.000,14.019){2}{\rule{0.800pt}{0.718pt}}
\multiput(271.40,262.00)(0.514,0.710){13}{\rule{0.124pt}{1.320pt}}
\multiput(268.34,262.00)(10.000,11.260){2}{\rule{0.800pt}{0.660pt}}
\multiput(281.40,276.00)(0.512,0.639){15}{\rule{0.123pt}{1.218pt}}
\multiput(278.34,276.00)(11.000,11.472){2}{\rule{0.800pt}{0.609pt}}
\multiput(292.40,290.00)(0.514,0.654){13}{\rule{0.124pt}{1.240pt}}
\multiput(289.34,290.00)(10.000,10.426){2}{\rule{0.800pt}{0.620pt}}
\multiput(302.40,303.00)(0.514,0.543){13}{\rule{0.124pt}{1.080pt}}
\multiput(299.34,303.00)(10.000,8.758){2}{\rule{0.800pt}{0.540pt}}
\multiput(312.40,314.00)(0.512,0.539){15}{\rule{0.123pt}{1.073pt}}
\multiput(309.34,314.00)(11.000,9.774){2}{\rule{0.800pt}{0.536pt}}
\multiput(322.00,327.40)(0.487,0.514){13}{\rule{1.000pt}{0.124pt}}
\multiput(322.00,324.34)(7.924,10.000){2}{\rule{0.500pt}{0.800pt}}
\multiput(332.00,337.40)(0.543,0.514){13}{\rule{1.080pt}{0.124pt}}
\multiput(332.00,334.34)(8.758,10.000){2}{\rule{0.540pt}{0.800pt}}
\multiput(343.00,347.40)(0.487,0.514){13}{\rule{1.000pt}{0.124pt}}
\multiput(343.00,344.34)(7.924,10.000){2}{\rule{0.500pt}{0.800pt}}
\multiput(353.00,357.40)(0.548,0.516){11}{\rule{1.089pt}{0.124pt}}
\multiput(353.00,354.34)(7.740,9.000){2}{\rule{0.544pt}{0.800pt}}
\multiput(363.00,366.40)(0.611,0.516){11}{\rule{1.178pt}{0.124pt}}
\multiput(363.00,363.34)(8.555,9.000){2}{\rule{0.589pt}{0.800pt}}
\multiput(374.00,375.40)(0.548,0.516){11}{\rule{1.089pt}{0.124pt}}
\multiput(374.00,372.34)(7.740,9.000){2}{\rule{0.544pt}{0.800pt}}
\multiput(384.00,384.40)(0.627,0.520){9}{\rule{1.200pt}{0.125pt}}
\multiput(384.00,381.34)(7.509,8.000){2}{\rule{0.600pt}{0.800pt}}
\multiput(394.00,392.40)(0.611,0.516){11}{\rule{1.178pt}{0.124pt}}
\multiput(394.00,389.34)(8.555,9.000){2}{\rule{0.589pt}{0.800pt}}
\multiput(405.00,401.40)(0.627,0.520){9}{\rule{1.200pt}{0.125pt}}
\multiput(405.00,398.34)(7.509,8.000){2}{\rule{0.600pt}{0.800pt}}
\multiput(415.00,409.40)(0.825,0.526){7}{\rule{1.457pt}{0.127pt}}
\multiput(415.00,406.34)(7.976,7.000){2}{\rule{0.729pt}{0.800pt}}
\multiput(426.00,416.40)(0.627,0.520){9}{\rule{1.200pt}{0.125pt}}
\multiput(426.00,413.34)(7.509,8.000){2}{\rule{0.600pt}{0.800pt}}
\multiput(436.00,424.40)(0.738,0.526){7}{\rule{1.343pt}{0.127pt}}
\multiput(436.00,421.34)(7.213,7.000){2}{\rule{0.671pt}{0.800pt}}
\multiput(446.00,431.40)(0.825,0.526){7}{\rule{1.457pt}{0.127pt}}
\multiput(446.00,428.34)(7.976,7.000){2}{\rule{0.729pt}{0.800pt}}
\multiput(457.00,438.40)(0.627,0.520){9}{\rule{1.200pt}{0.125pt}}
\multiput(457.00,435.34)(7.509,8.000){2}{\rule{0.600pt}{0.800pt}}
\multiput(467.00,446.39)(1.020,0.536){5}{\rule{1.667pt}{0.129pt}}
\multiput(467.00,443.34)(7.541,6.000){2}{\rule{0.833pt}{0.800pt}}
\multiput(478.00,452.40)(0.738,0.526){7}{\rule{1.343pt}{0.127pt}}
\multiput(478.00,449.34)(7.213,7.000){2}{\rule{0.671pt}{0.800pt}}
\multiput(488.00,459.40)(0.738,0.526){7}{\rule{1.343pt}{0.127pt}}
\multiput(488.00,456.34)(7.213,7.000){2}{\rule{0.671pt}{0.800pt}}
\multiput(498.00,466.39)(1.020,0.536){5}{\rule{1.667pt}{0.129pt}}
\multiput(498.00,463.34)(7.541,6.000){2}{\rule{0.833pt}{0.800pt}}
\multiput(509.00,472.40)(0.738,0.526){7}{\rule{1.343pt}{0.127pt}}
\multiput(509.00,469.34)(7.213,7.000){2}{\rule{0.671pt}{0.800pt}}
\multiput(519.00,479.39)(1.020,0.536){5}{\rule{1.667pt}{0.129pt}}
\multiput(519.00,476.34)(7.541,6.000){2}{\rule{0.833pt}{0.800pt}}
\multiput(530.00,485.39)(0.909,0.536){5}{\rule{1.533pt}{0.129pt}}
\multiput(530.00,482.34)(6.817,6.000){2}{\rule{0.767pt}{0.800pt}}
\multiput(540.00,491.39)(0.909,0.536){5}{\rule{1.533pt}{0.129pt}}
\multiput(540.00,488.34)(6.817,6.000){2}{\rule{0.767pt}{0.800pt}}
\multiput(550.00,497.39)(1.020,0.536){5}{\rule{1.667pt}{0.129pt}}
\multiput(550.00,494.34)(7.541,6.000){2}{\rule{0.833pt}{0.800pt}}
\multiput(561.00,503.39)(0.909,0.536){5}{\rule{1.533pt}{0.129pt}}
\multiput(561.00,500.34)(6.817,6.000){2}{\rule{0.767pt}{0.800pt}}
\multiput(571.00,509.39)(1.020,0.536){5}{\rule{1.667pt}{0.129pt}}
\multiput(571.00,506.34)(7.541,6.000){2}{\rule{0.833pt}{0.800pt}}
\multiput(582.00,515.39)(0.909,0.536){5}{\rule{1.533pt}{0.129pt}}
\multiput(582.00,512.34)(6.817,6.000){2}{\rule{0.767pt}{0.800pt}}
\multiput(592.00,521.38)(1.264,0.560){3}{\rule{1.800pt}{0.135pt}}
\multiput(592.00,518.34)(6.264,5.000){2}{\rule{0.900pt}{0.800pt}}
\multiput(602.00,526.39)(1.020,0.536){5}{\rule{1.667pt}{0.129pt}}
\multiput(602.00,523.34)(7.541,6.000){2}{\rule{0.833pt}{0.800pt}}
\multiput(613.00,532.38)(1.264,0.560){3}{\rule{1.800pt}{0.135pt}}
\multiput(613.00,529.34)(6.264,5.000){2}{\rule{0.900pt}{0.800pt}}
\multiput(623.00,537.39)(1.020,0.536){5}{\rule{1.667pt}{0.129pt}}
\multiput(623.00,534.34)(7.541,6.000){2}{\rule{0.833pt}{0.800pt}}
\multiput(634.00,543.38)(1.264,0.560){3}{\rule{1.800pt}{0.135pt}}
\multiput(634.00,540.34)(6.264,5.000){2}{\rule{0.900pt}{0.800pt}}
\multiput(644.00,548.38)(1.264,0.560){3}{\rule{1.800pt}{0.135pt}}
\multiput(644.00,545.34)(6.264,5.000){2}{\rule{0.900pt}{0.800pt}}
\multiput(654.00,553.39)(1.020,0.536){5}{\rule{1.667pt}{0.129pt}}
\multiput(654.00,550.34)(7.541,6.000){2}{\rule{0.833pt}{0.800pt}}
\multiput(665.00,559.38)(1.264,0.560){3}{\rule{1.800pt}{0.135pt}}
\multiput(665.00,556.34)(6.264,5.000){2}{\rule{0.900pt}{0.800pt}}
\multiput(675.00,564.38)(1.432,0.560){3}{\rule{1.960pt}{0.135pt}}
\multiput(675.00,561.34)(6.932,5.000){2}{\rule{0.980pt}{0.800pt}}
\multiput(686.00,569.38)(1.264,0.560){3}{\rule{1.800pt}{0.135pt}}
\multiput(686.00,566.34)(6.264,5.000){2}{\rule{0.900pt}{0.800pt}}
\multiput(696.00,574.38)(1.264,0.560){3}{\rule{1.800pt}{0.135pt}}
\multiput(696.00,571.34)(6.264,5.000){2}{\rule{0.900pt}{0.800pt}}
\multiput(706.00,579.38)(1.432,0.560){3}{\rule{1.960pt}{0.135pt}}
\multiput(706.00,576.34)(6.932,5.000){2}{\rule{0.980pt}{0.800pt}}
\multiput(717.00,584.38)(1.264,0.560){3}{\rule{1.800pt}{0.135pt}}
\multiput(717.00,581.34)(6.264,5.000){2}{\rule{0.900pt}{0.800pt}}
\multiput(727.00,589.38)(1.432,0.560){3}{\rule{1.960pt}{0.135pt}}
\multiput(727.00,586.34)(6.932,5.000){2}{\rule{0.980pt}{0.800pt}}
\put(738,593.34){\rule{2.200pt}{0.800pt}}
\multiput(738.00,591.34)(5.434,4.000){2}{\rule{1.100pt}{0.800pt}}
\multiput(748.00,598.38)(1.264,0.560){3}{\rule{1.800pt}{0.135pt}}
\multiput(748.00,595.34)(6.264,5.000){2}{\rule{0.900pt}{0.800pt}}
\multiput(758.00,603.38)(1.432,0.560){3}{\rule{1.960pt}{0.135pt}}
\multiput(758.00,600.34)(6.932,5.000){2}{\rule{0.980pt}{0.800pt}}
\multiput(769.00,608.38)(1.264,0.560){3}{\rule{1.800pt}{0.135pt}}
\multiput(769.00,605.34)(6.264,5.000){2}{\rule{0.900pt}{0.800pt}}
\put(779,612.34){\rule{2.400pt}{0.800pt}}
\multiput(779.00,610.34)(6.019,4.000){2}{\rule{1.200pt}{0.800pt}}
\multiput(790.00,617.38)(1.264,0.560){3}{\rule{1.800pt}{0.135pt}}
\multiput(790.00,614.34)(6.264,5.000){2}{\rule{0.900pt}{0.800pt}}
\put(800,621.34){\rule{2.200pt}{0.800pt}}
\multiput(800.00,619.34)(5.434,4.000){2}{\rule{1.100pt}{0.800pt}}
\multiput(810.00,626.38)(1.432,0.560){3}{\rule{1.960pt}{0.135pt}}
\multiput(810.00,623.34)(6.932,5.000){2}{\rule{0.980pt}{0.800pt}}
\put(821,630.34){\rule{2.200pt}{0.800pt}}
\multiput(821.00,628.34)(5.434,4.000){2}{\rule{1.100pt}{0.800pt}}
\multiput(831.00,635.38)(1.432,0.560){3}{\rule{1.960pt}{0.135pt}}
\multiput(831.00,632.34)(6.932,5.000){2}{\rule{0.980pt}{0.800pt}}
\put(842,639.34){\rule{2.200pt}{0.800pt}}
\multiput(842.00,637.34)(5.434,4.000){2}{\rule{1.100pt}{0.800pt}}
\put(852,643.34){\rule{2.200pt}{0.800pt}}
\multiput(852.00,641.34)(5.434,4.000){2}{\rule{1.100pt}{0.800pt}}
\multiput(862.00,648.38)(1.432,0.560){3}{\rule{1.960pt}{0.135pt}}
\multiput(862.00,645.34)(6.932,5.000){2}{\rule{0.980pt}{0.800pt}}
\put(873,652.34){\rule{2.200pt}{0.800pt}}
\multiput(873.00,650.34)(5.434,4.000){2}{\rule{1.100pt}{0.800pt}}
\put(883,656.34){\rule{2.400pt}{0.800pt}}
\multiput(883.00,654.34)(6.019,4.000){2}{\rule{1.200pt}{0.800pt}}
\put(894,660.34){\rule{2.200pt}{0.800pt}}
\multiput(894.00,658.34)(5.434,4.000){2}{\rule{1.100pt}{0.800pt}}
\multiput(904.00,665.38)(1.264,0.560){3}{\rule{1.800pt}{0.135pt}}
\multiput(904.00,662.34)(6.264,5.000){2}{\rule{0.900pt}{0.800pt}}
\put(914,669.34){\rule{2.400pt}{0.800pt}}
\multiput(914.00,667.34)(6.019,4.000){2}{\rule{1.200pt}{0.800pt}}
\put(925,673.34){\rule{2.200pt}{0.800pt}}
\multiput(925.00,671.34)(5.434,4.000){2}{\rule{1.100pt}{0.800pt}}
\put(935,677.34){\rule{2.400pt}{0.800pt}}
\multiput(935.00,675.34)(6.019,4.000){2}{\rule{1.200pt}{0.800pt}}
\put(946,681.34){\rule{2.200pt}{0.800pt}}
\multiput(946.00,679.34)(5.434,4.000){2}{\rule{1.100pt}{0.800pt}}
\put(956,685.34){\rule{2.200pt}{0.800pt}}
\multiput(956.00,683.34)(5.434,4.000){2}{\rule{1.100pt}{0.800pt}}
\put(966,689.34){\rule{2.400pt}{0.800pt}}
\multiput(966.00,687.34)(6.019,4.000){2}{\rule{1.200pt}{0.800pt}}
\put(977,693.34){\rule{2.200pt}{0.800pt}}
\multiput(977.00,691.34)(5.434,4.000){2}{\rule{1.100pt}{0.800pt}}
\put(987,697.34){\rule{2.400pt}{0.800pt}}
\multiput(987.00,695.34)(6.019,4.000){2}{\rule{1.200pt}{0.800pt}}
\put(998,701.34){\rule{2.200pt}{0.800pt}}
\multiput(998.00,699.34)(5.434,4.000){2}{\rule{1.100pt}{0.800pt}}
\put(1008,705.34){\rule{2.200pt}{0.800pt}}
\multiput(1008.00,703.34)(5.434,4.000){2}{\rule{1.100pt}{0.800pt}}
\put(1018,709.34){\rule{2.400pt}{0.800pt}}
\multiput(1018.00,707.34)(6.019,4.000){2}{\rule{1.200pt}{0.800pt}}
\put(1029,712.84){\rule{2.409pt}{0.800pt}}
\multiput(1029.00,711.34)(5.000,3.000){2}{\rule{1.204pt}{0.800pt}}
\put(1039,716.34){\rule{2.400pt}{0.800pt}}
\multiput(1039.00,714.34)(6.019,4.000){2}{\rule{1.200pt}{0.800pt}}
\put(1050,720.34){\rule{2.200pt}{0.800pt}}
\multiput(1050.00,718.34)(5.434,4.000){2}{\rule{1.100pt}{0.800pt}}
\put(1060,724.34){\rule{2.200pt}{0.800pt}}
\multiput(1060.00,722.34)(5.434,4.000){2}{\rule{1.100pt}{0.800pt}}
\put(1070,728.34){\rule{2.400pt}{0.800pt}}
\multiput(1070.00,726.34)(6.019,4.000){2}{\rule{1.200pt}{0.800pt}}
\put(1081,731.84){\rule{2.409pt}{0.800pt}}
\multiput(1081.00,730.34)(5.000,3.000){2}{\rule{1.204pt}{0.800pt}}
\put(1091,735.34){\rule{2.400pt}{0.800pt}}
\multiput(1091.00,733.34)(6.019,4.000){2}{\rule{1.200pt}{0.800pt}}
\put(1102,739.34){\rule{2.200pt}{0.800pt}}
\multiput(1102.00,737.34)(5.434,4.000){2}{\rule{1.100pt}{0.800pt}}
\put(1112,742.84){\rule{2.409pt}{0.800pt}}
\multiput(1112.00,741.34)(5.000,3.000){2}{\rule{1.204pt}{0.800pt}}
\put(1122,746.34){\rule{2.400pt}{0.800pt}}
\multiput(1122.00,744.34)(6.019,4.000){2}{\rule{1.200pt}{0.800pt}}
\put(1133,750.34){\rule{2.200pt}{0.800pt}}
\multiput(1133.00,748.34)(5.434,4.000){2}{\rule{1.100pt}{0.800pt}}
\put(1143,753.84){\rule{2.650pt}{0.800pt}}
\multiput(1143.00,752.34)(5.500,3.000){2}{\rule{1.325pt}{0.800pt}}
\put(1154,757.34){\rule{2.200pt}{0.800pt}}
\multiput(1154.00,755.34)(5.434,4.000){2}{\rule{1.100pt}{0.800pt}}
\put(1164,760.84){\rule{2.409pt}{0.800pt}}
\multiput(1164.00,759.34)(5.000,3.000){2}{\rule{1.204pt}{0.800pt}}
\put(1174,764.34){\rule{2.400pt}{0.800pt}}
\multiput(1174.00,762.34)(6.019,4.000){2}{\rule{1.200pt}{0.800pt}}
\put(1185,767.84){\rule{2.409pt}{0.800pt}}
\multiput(1185.00,766.34)(5.000,3.000){2}{\rule{1.204pt}{0.800pt}}
\put(1195,771.34){\rule{2.400pt}{0.800pt}}
\multiput(1195.00,769.34)(6.019,4.000){2}{\rule{1.200pt}{0.800pt}}
\put(1206,774.84){\rule{2.409pt}{0.800pt}}
\multiput(1206.00,773.34)(5.000,3.000){2}{\rule{1.204pt}{0.800pt}}
\put(1216,778.34){\rule{2.200pt}{0.800pt}}
\multiput(1216.00,776.34)(5.434,4.000){2}{\rule{1.100pt}{0.800pt}}
\put(1226,781.84){\rule{2.650pt}{0.800pt}}
\multiput(1226.00,780.34)(5.500,3.000){2}{\rule{1.325pt}{0.800pt}}
\put(1237,785.34){\rule{2.200pt}{0.800pt}}
\multiput(1237.00,783.34)(5.434,4.000){2}{\rule{1.100pt}{0.800pt}}
\put(1247,788.84){\rule{2.650pt}{0.800pt}}
\multiput(1247.00,787.34)(5.500,3.000){2}{\rule{1.325pt}{0.800pt}}
\sbox{\plotpoint}{\rule[-0.500pt]{1.000pt}{1.000pt}}%
\put(228,155){\usebox{\plotpoint}}
\multiput(228,155)(9.282,18.564){2}{\usebox{\plotpoint}}
\put(247.96,191.34){\usebox{\plotpoint}}
\multiput(249,193)(12.655,16.451){0}{\usebox{\plotpoint}}
\put(260.62,207.77){\usebox{\plotpoint}}
\put(274.63,223.09){\usebox{\plotpoint}}
\put(289.88,237.08){\usebox{\plotpoint}}
\multiput(291,238)(15.427,13.885){0}{\usebox{\plotpoint}}
\put(305.57,250.66){\usebox{\plotpoint}}
\multiput(311,255)(16.786,12.208){0}{\usebox{\plotpoint}}
\put(322.16,263.13){\usebox{\plotpoint}}
\put(338.88,275.38){\usebox{\plotpoint}}
\multiput(343,278)(17.004,11.902){0}{\usebox{\plotpoint}}
\put(356.14,286.88){\usebox{\plotpoint}}
\put(373.76,297.85){\usebox{\plotpoint}}
\multiput(374,298)(17.798,10.679){0}{\usebox{\plotpoint}}
\put(391.56,308.53){\usebox{\plotpoint}}
\multiput(394,310)(18.221,9.939){0}{\usebox{\plotpoint}}
\put(409.81,318.40){\usebox{\plotpoint}}
\multiput(415,321)(18.221,9.939){0}{\usebox{\plotpoint}}
\put(428.17,328.08){\usebox{\plotpoint}}
\multiput(436,332)(18.564,9.282){0}{\usebox{\plotpoint}}
\put(446.74,337.34){\usebox{\plotpoint}}
\put(465.49,346.24){\usebox{\plotpoint}}
\multiput(467,347)(18.895,8.589){0}{\usebox{\plotpoint}}
\put(484.24,355.12){\usebox{\plotpoint}}
\multiput(488,357)(18.564,9.282){0}{\usebox{\plotpoint}}
\put(503.05,363.84){\usebox{\plotpoint}}
\multiput(509,366)(18.564,9.282){0}{\usebox{\plotpoint}}
\put(522.05,372.11){\usebox{\plotpoint}}
\multiput(530,375)(18.564,9.282){0}{\usebox{\plotpoint}}
\put(541.04,380.41){\usebox{\plotpoint}}
\put(560.43,387.79){\usebox{\plotpoint}}
\multiput(561,388)(19.271,7.708){0}{\usebox{\plotpoint}}
\put(579.54,395.88){\usebox{\plotpoint}}
\multiput(582,397)(19.271,7.708){0}{\usebox{\plotpoint}}
\put(598.76,403.71){\usebox{\plotpoint}}
\multiput(602,405)(20.024,5.461){0}{\usebox{\plotpoint}}
\put(618.45,410.18){\usebox{\plotpoint}}
\multiput(623,412)(19.506,7.093){0}{\usebox{\plotpoint}}
\put(637.85,417.54){\usebox{\plotpoint}}
\multiput(644,420)(19.271,7.708){0}{\usebox{\plotpoint}}
\put(657.24,424.88){\usebox{\plotpoint}}
\multiput(665,427)(19.271,7.708){0}{\usebox{\plotpoint}}
\put(676.83,431.67){\usebox{\plotpoint}}
\multiput(686,435)(19.880,5.964){0}{\usebox{\plotpoint}}
\put(696.52,438.21){\usebox{\plotpoint}}
\put(716.17,444.77){\usebox{\plotpoint}}
\multiput(717,445)(19.271,7.708){0}{\usebox{\plotpoint}}
\put(735.80,451.40){\usebox{\plotpoint}}
\multiput(738,452)(19.271,7.708){0}{\usebox{\plotpoint}}
\put(755.38,458.22){\usebox{\plotpoint}}
\multiput(758,459)(20.024,5.461){0}{\usebox{\plotpoint}}
\put(775.15,464.46){\usebox{\plotpoint}}
\multiput(779,466)(20.024,5.461){0}{\usebox{\plotpoint}}
\put(794.99,470.50){\usebox{\plotpoint}}
\multiput(800,472)(19.880,5.964){0}{\usebox{\plotpoint}}
\put(814.90,476.34){\usebox{\plotpoint}}
\multiput(821,478)(19.271,7.708){0}{\usebox{\plotpoint}}
\put(834.53,482.96){\usebox{\plotpoint}}
\multiput(842,485)(19.880,5.964){0}{\usebox{\plotpoint}}
\put(854.47,488.74){\usebox{\plotpoint}}
\multiput(862,491)(20.024,5.461){0}{\usebox{\plotpoint}}
\put(874.43,494.43){\usebox{\plotpoint}}
\multiput(883,497)(20.024,5.461){0}{\usebox{\plotpoint}}
\put(894.39,500.12){\usebox{\plotpoint}}
\multiput(904,503)(19.880,5.964){0}{\usebox{\plotpoint}}
\put(914.27,506.07){\usebox{\plotpoint}}
\put(934.23,511.77){\usebox{\plotpoint}}
\multiput(935,512)(20.024,5.461){0}{\usebox{\plotpoint}}
\put(954.38,516.68){\usebox{\plotpoint}}
\multiput(956,517)(19.880,5.964){0}{\usebox{\plotpoint}}
\put(974.36,522.28){\usebox{\plotpoint}}
\multiput(977,523)(19.880,5.964){0}{\usebox{\plotpoint}}
\put(994.31,527.99){\usebox{\plotpoint}}
\multiput(998,529)(20.352,4.070){0}{\usebox{\plotpoint}}
\put(1014.45,532.93){\usebox{\plotpoint}}
\multiput(1018,534)(20.024,5.461){0}{\usebox{\plotpoint}}
\put(1034.41,538.62){\usebox{\plotpoint}}
\multiput(1039,540)(20.421,3.713){0}{\usebox{\plotpoint}}
\put(1054.58,543.37){\usebox{\plotpoint}}
\multiput(1060,545)(19.880,5.964){0}{\usebox{\plotpoint}}
\put(1074.58,548.83){\usebox{\plotpoint}}
\multiput(1081,550)(19.880,5.964){0}{\usebox{\plotpoint}}
\put(1094.66,554.00){\usebox{\plotpoint}}
\multiput(1102,556)(20.352,4.070){0}{\usebox{\plotpoint}}
\put(1114.82,558.85){\usebox{\plotpoint}}
\multiput(1122,561)(20.421,3.713){0}{\usebox{\plotpoint}}
\put(1134.99,563.60){\usebox{\plotpoint}}
\multiput(1143,566)(20.024,5.461){0}{\usebox{\plotpoint}}
\put(1154.98,569.20){\usebox{\plotpoint}}
\multiput(1164,571)(19.880,5.964){0}{\usebox{\plotpoint}}
\put(1175.09,574.20){\usebox{\plotpoint}}
\multiput(1185,576)(19.880,5.964){0}{\usebox{\plotpoint}}
\put(1195.24,579.04){\usebox{\plotpoint}}
\put(1215.63,582.93){\usebox{\plotpoint}}
\multiput(1216,583)(19.880,5.964){0}{\usebox{\plotpoint}}
\put(1235.78,587.78){\usebox{\plotpoint}}
\multiput(1237,588)(19.880,5.964){0}{\usebox{\plotpoint}}
\put(1255.93,592.62){\usebox{\plotpoint}}
\put(1258,593){\usebox{\plotpoint}}
\end{picture}

{\footnotesize
\caption{Average ratio between the time spent to communicate data for the
systolic algorithm against the hyper-systolic one for the example of
the matrix-matrix product.
The squares are obtained using the best basis, the crosses using the
regular basis.
The lines represent the ideal cases for the best basis (solid) and
regular basis (dashed), respectively. The errorbars of the average
over the different matrix dimensions are smaller than the plotted
symbols.}}
\end{figure}
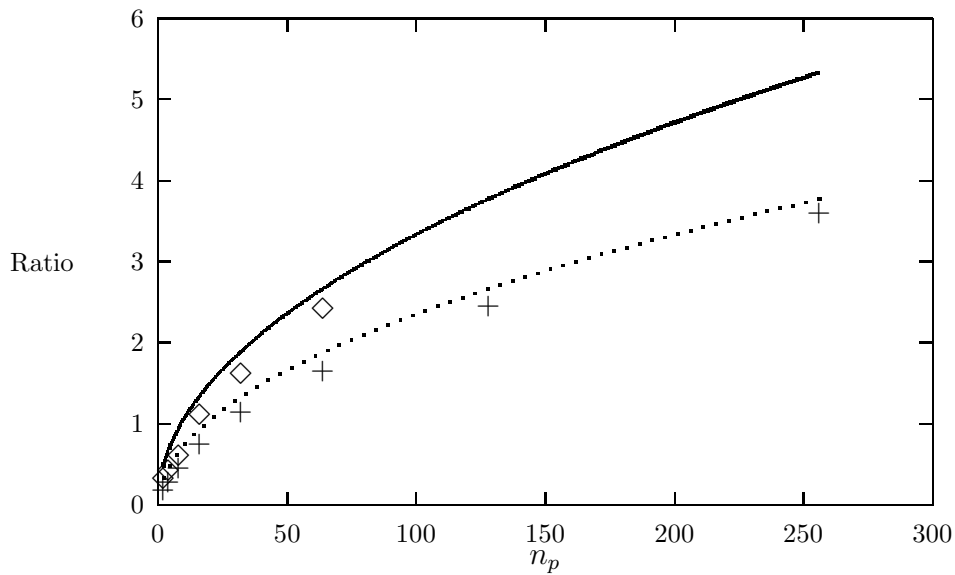

\begin{thebibliography}{99}
\bibitem{schilling} Th. Lippert, A. Seyfried, A. Bode and K.
Schilling, WUB 95-13, HLRZ 32/95, hep-lat/9507021
\bibitem{sys} H.T.Kung, Computer Vol. 15 (1982) 37\\
W.Smith, Comput. Phys. Commun. 62 (1991) 229
\end{thebibliography}
\end{document}